\documentclass[noshowpacs,amsmath,twocolumn,
aps,prb]
{revtex4-1}

\usepackage{setspace}
\usepackage{amsmath}
\usepackage{graphicx}
\usepackage[nearskip,margin = 0pt]{subfig}
\usepackage{subfig}
\usepackage{verbatim}
\usepackage{amsfonts}
\usepackage{amssymb}
\usepackage{epstopdf} 
\usepackage{ulem}
\usepackage{xcolor}
\DeclareGraphicsExtensions{.pdf,.eps,.png,.jpg,.mps}
\usepackage{ragged2e}
\usepackage{hyperref}
\newcommand{\FP}{Fabry–P\'erot }
\hypersetup{
    colorlinks=true,
    linkcolor=blue,
    citecolor=blue,
    filecolor=blue,      
    urlcolor=blue,
}

\begin{document}

\title{Dispersive-wave-agile optical frequency division}

\author{Qing-Xin Ji$^{1}$, Wei Zhang$^{2,\dagger}$, Peng Liu$^{1}$, Warren Jin$^{3,4}$, Joel Guo$^{2}$, Jonathan Peters$^{2}$, Lue Wu$^{1}$, Avi Feshali$^{4}$, Mario Paniccia$^{4}$,  Vladimir Ilchenko$^{2}$, John Bowers$^{3}$, Andrey Matsko$^{2}$, and Kerry Vahala$^{1,\dagger}$\\
$^1$T. J. Watson Laboratory of Applied Physics, California Institute of Technology, Pasadena, CA 91125, USA.\\
$^2$Jet Propulsion Laboratory, California Institute of Technology, Pasadena, CA 91125, USA. \\
$^3$ECE Department, University of California Santa Barbara, Santa Barbara, CA 93106, USA.\\
$^4$Anello Photonics, Santa Clara, CA 95054, USA.\\
$^\dagger$Corresponding authors: wei.zhang@jpl.nasa.gov, vahala@caltech.edu}

\maketitle

{\bf The remarkable frequency stability of resonant systems in the optical domain (optical cavities and atomic transitions) can be harnessed at frequency scales accessible by electronics using optical frequency division. This capability is revolutionizing technologies spanning time keeping to high-performance electrical signal sources.  A version of the technique called 2-point optical frequency division (2P-OFD) is proving advantageous for application to high-performance signal sources. In 2P-OFD, an optical cavity anchors two spectral endpoints defined by lines of a frequency comb. The comb need not be self-referenced, which greatly simplifies the system architecture and reduces power requirements. Here, a 2P-OFD microwave signal source is demonstrated with record-low phase noise using a microcomb. Key to this advance is a spectral endpoint defined by a frequency agile single-mode dispersive wave that is emitted by the microcomb soliton. Moreover, the system frequency reference is a compact all-solid-state optical cavity with a record $Q$-factor. The results advance integrable microcomb-based signal sources into the performance realm of much larger microwave sources. }


\section{Introduction}
An octave spanning frequency comb is required to perform conventional optical frequency division, wherein the maximum frequency division ratio is attained as, for example, is required to count optical cycles \cite{diddams2020optical}. This method has led to the most stable microwave signals ever generated \cite{fortier2011generation,xie2017photonic,nakamura2020coherent}. Two-point optical frequency division (2P-OFD) trades-off a portion of this performance for technical simplification resulting from using a non-self-referenced (narrower spectral span) frequency comb \cite{swann2011microwave,li2014electro}. This tradeoff has made possible compact commercial microwave systems based on electro-optically generated combs \cite{li2023small} and chip-integrable microwave signal sources using microcombs \cite{kudelin2023photonic, sun2023integrated,zhao2023all,kwon2022ultrastable}. In microcomb-based 2P-OFD, spectral endpoints of the comb are locked to two frequencies of a reference cavity. This transfers relative frequency stability from the cavity to the comb repetition rate, which can be photodetected to generate a microwave signal. Control of power and frequency at the comb spectral endpoints is critically important to 2P-OFD. High power at these endpoints is necessary for low phase noise microwave signal generation, and endpoint frequency control is essential to roughly align these frequencies with reference laser lines. This combination of features is challenging in bright soliton microcombs, which also offer the highest frequency division performance through their broad spectral reach. Here, 2P-OFD is demonstrated using a bright soliton microcomb with spectral endpoints that are both high in power and frequency tunable. These features in combination with the spectral reach of microcomb achieve optical division noise reduction to a detectable microwave signal that is 30X greater than in recent reports \cite{kudelin2023photonic}. The microcomb is referenced to a solid state cavity with 8-fold higher Q factor relative to earlier designs \cite{zhang2020ultranarrow}, and together make possible a record-low phase noise for chip-integrable components. 

\begin{figure*}[ht!]
\begin{centering}
\includegraphics[width=\linewidth]{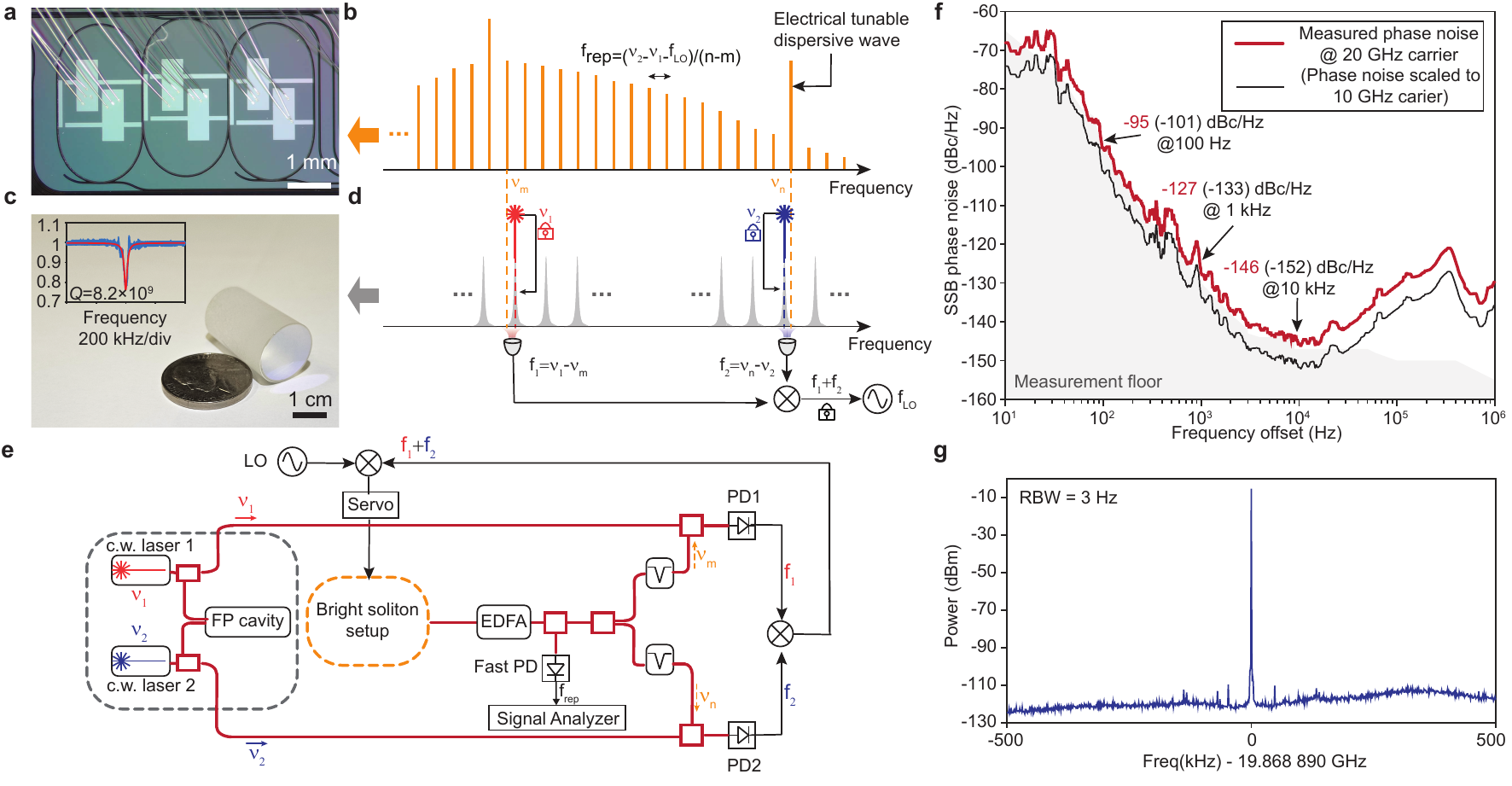}
\captionsetup{singlelinecheck=off, justification = RaggedLeft}
\caption{{\bf The low-noise, miniaturized optical frequency division architecture. }
\textbf{a}, Photograph of the three-coupled-ring (3CR) resonator with heaters for differential temperature tuning of the three rings. 
\textbf{b}, Schematic of the microcomb in the 2P-OFD system. A dispersive wave at frequency $\nu_{n}$ is electrically tuned to closely match the frequency $\nu_{2}$ of the c.w. laser. 
\textbf{c}, Photograph of the miniature, vacuum-free \FP (FP) cavity with a US nickel. 
Inset: \FP cavity reflection spectrum showing Q factor of 8.2 billion. 
\textbf{d}, Schematic of the FP cavity spectrum in the 2P-OFD system, wherein two c.w. lasers at frequencies $\nu_1$ and $\nu_2$ (separated by approximately 3 THz) are locked to the cavity. Beatnote frequencies f$_1$ and f$_2$ between comb lines and the nearby c.w. lasers are generated via photo-detection ($\sim$1 GHz), and mixed to generate their sum frequency. The summed frequency is stabilized to a local oscillator ($f_{\rm LO}$) by feedback to the microcomb. The microcomb repetition rate is given by $f_{\rm rep}=(\nu_2-\nu_1-f_{\rm LO})/(n-m)$, which divides the 3 THz laser separation down to the microcomb repetition rate at microwave frequency (20 GHz). 
\textbf{e}, Detailed experimental setup of the 2P-OFD schematic in panel d. 
EDFA: Erbium-doped fiber amplifier. PD: photo detector. 
LO: local oscillator. 
\textbf{f}, Single-sideband phase noise spectrum measurement with carrier frequency near 20 GHz (red). The instrumental measurement floor is the gray shaded area. 
For comparison, the phase noise is scaled by $\mathcal{L}_{\phi,10 \mathrm{GHz}}=\mathcal{L}_{\phi,f_{\rm rep}} - 20\times \mathrm{log}_{10}(f_{\rm rep}/10~\mathrm{GHz})$, and plotted in black. 
\textbf{g}, High-spectral purity microwave tone near 20 GHz generated by the 2P-OFD system. The resolution bandwidth of the measured microwave tone is 3 Hz.
}
\label{Fig1}
\end{centering}
\end{figure*}

High power spectral endpoints with tunable frequency control are fully achieved in electro-optic combs \cite{li2014electro} on account of their tunable repetition rate and ability to efficiently throw pump power to the comb spectral wings. High power spectral endpoints (without tuning control) is also a feature of normal dispersion microcombs \cite{ji2023integrated} (sometimes called dark pulse combs \cite{xue2015normal}) and for this reason these combs have been used recently for microwave signal generation \cite{kudelin2023photonic}. There, the lack of endpoint tuning is less important when the comb repetition rate is low enough to guarantee comb line spectral proximity to the optical reference. However, both of the these comb generation methods tend to provide narrower comb spectral spans, thereby limiting the amount of optical frequency division. And while bright soliton microcombs offer broader spectral spans, their per line comb power is limited on account of the steep (exponential) roll-off in comb spectral envelope away from the pump line \cite{kippenberg2018dissipative}. 

Dispersive wave generation provides a solution to this problem \cite{brasch2017self} and has been implemented in optical synthesizer \cite{spencer2018optical} and optical clock \cite{newman2019architecture} demonstrations featuring octave-span microcombs and full optical frequency division. Dispersive waves form when soliton comb lines phase match to resonator modes (from either the same or another mode family). They appear as a spectrally-local enhancement in comb line power near the phase matching frequency. Strong single-line dispersive waves can form in cases where a single cavity mode phase matches to the soliton \cite{yi2017single}, however geometrical control of dispersion to accomplish this matching is challenging. Tuning of this phase matching condition is possible in coupled ring resonators \cite{okawachi2022active}. In this paper, broad tuning of single line dispersive waves is used to implement broadband 2P-OFD at directly detectable microwave repetition rates.

\begin{figure*}[t!]
\begin{centering}
\includegraphics[width=\linewidth]{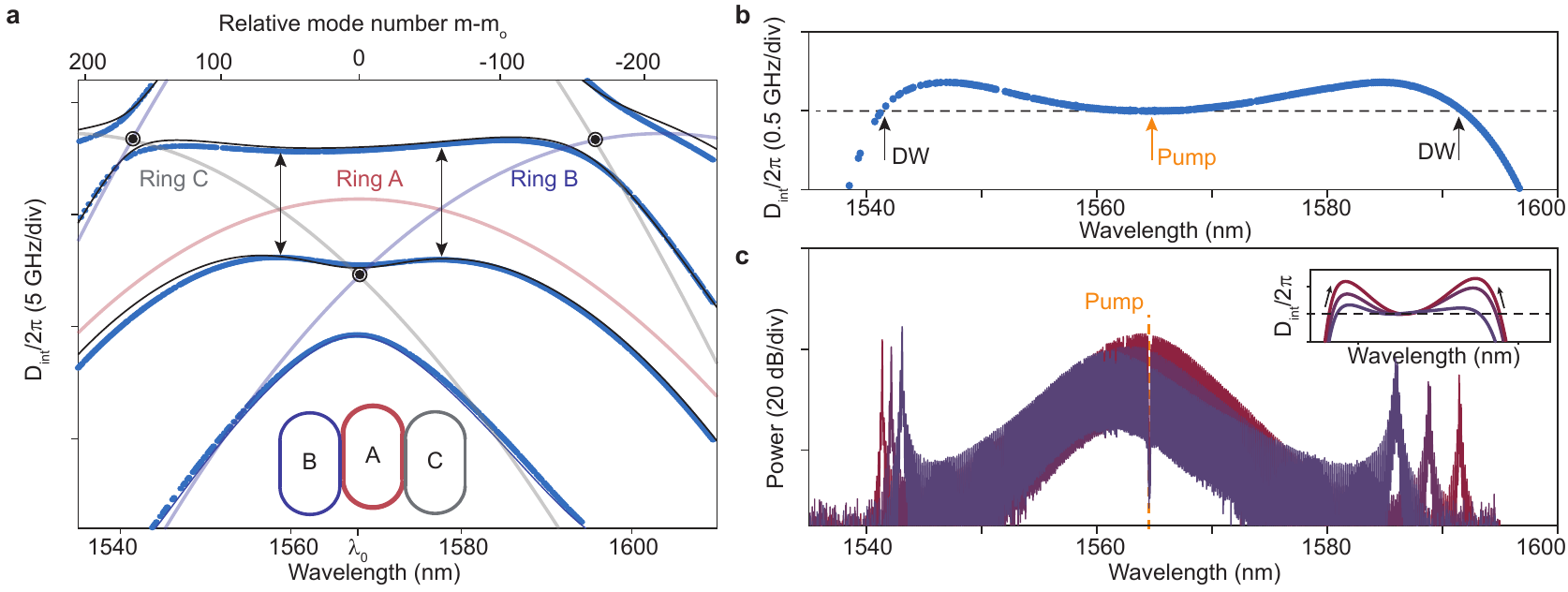}
\captionsetup{singlelinecheck=off, justification = RaggedRight}
\caption{{\bf Dispersion and bright soliton microcomb with electrically tunable dispersive wave.}
\textbf{a}, Measured dispersion spectrum showing the 3 dispersion bands of the 3CR device (blue dots). The theoretical fitting is plotted as the solid curves, and the inferred dispersion of the uncoupled rings are the shaded lines (red, blue and gray for ring A, B and C respectively). 
The relative mode number $m-m_{\rm o}$ is plotted in the upper axis, where $m_{\rm o}$ corresponds to the intersection of the uncoupled ring B and ring C dispersion curves. The corresponding wavelength is $\lambda_{\rm o}=1568$ nm (lower axis).
\textbf{b}, Isolated dispersion spectrum for the upper band in panel a. Dashed horizontal line gives location of the dispersive wave phase matching wavelengths.  
\textbf{c}, Optical spectral of bright soliton microcomb showing tuning of the dispersive waves. 
Inset: illustration of the dispersion changing when the heaters are differentially fine-tuned. 
The arrows indicate the change of the dispersion curves when ring A is heated.
}
\label{Fig2}
\end{centering}
\end{figure*}

\medskip

\noindent {\bf 2P-OFD measurement.}
The microcomb uses a 3-coupled-ring (3CR) design fabricated using ultra-low-loss silicon nitride (Fig. \ref{Fig1}a). The resonator modelocks through formation of femto-second pulse pairs \cite{yuan2023soliton}. As described below, differential heating of the rings allows both broad dispersion tuning to setup the pulse pair operation as well as fine tuning for control of dispersive wave emission. Further details about the 3CR properties are in the Methods. 
Two lines of the microcomb ($\nu_m$ near the pump frequency) and $\nu_n$ (the dispersive wave frequency) serve as spectral endpoints for 2P-OFD (Fig. \ref{Fig1}b). 
Besides the microcomb, two c.w. lasers are stabilized to a miniature high-finesse, vacuum-free FP resonator (Fig. \ref{Fig1}c) that provides reference frequencies $\nu_1$ and $\nu_2$, as in Fig. \ref{Fig1}d. 

The detailed experimental setup is depicted in Fig. \ref{Fig1}e. The microcomb is pumped at the bus waveguide coupled to the central ring in Fig. \ref{Fig1}a, and comb power is collected at the waveguide coupled to the left ring (drop port) and amplified by an Erbium-doped fiber amplifier to 60 mW.  75 \% of the amplified power is split evenly and filtered by two optical bandpass filters to select two desired comb lines. These are combined with their respective stabilized c.w. laser, and detected by two photo detectors (New Focus 1611) to generate beatnotes at frequencies f$_1$ and f$_2$. The formation of the dispersive wave improves the signal to noise ratio (SNR) of this beatnote by 30 dB (Extended Fig. \ref{ExFig4}). This enhancement makes possible 2P-OFD over this broader comb span of 3 THz (i.e., $\nu_{n}$- $\nu_{m}$ in Fig. \ref{Fig1}b). The two beatnote signals are electrically amplified before mixing to generate their frequency-summed signal.  
This signal at frequency f$_{\rm LO}$ is amplified and mixed with a local oscillator at 1.25 GHz to generate the error signal. The error signal is processed by a servo (Vescent D2-125) for feedback to the microcomb to control its repetition rate (Fig. \ref{Fig1}e). This closed loop thereby implements 2P-OFD. Additional details are provided in the caption to Fig. \ref{Fig1}d.

Details on the \FP cavity are provided elsewhere \cite{zhang2020ultranarrow} (and in Methods), but briefly it is a bulk fused silica rod with high-reflectivity coatings. 
The $FSR$ is 4.0 GHz and cavity $Q$ factor is as high as 8.2$\times 10^{9}$ (inset of Fig. \ref{Fig1}c). 
The $Q$ factor is improved by more than 8 times compared with a previous demonstration \cite{zhang2020ultranarrow}.
The vacuum-free nature of the FP cavity simplifies the operation compared with the vacuum-based reference cavities \cite{jin2022micro,guo2022chip,mclemore2022miniaturizing} and its ultra-high $Q$ factor boosts its performance in stabilizing lasers compared with other vacuum-free reference cavities \cite{ilchenko2006optical,liu202236,liu2023high}. 
Upon Pound-Drever-Hall locking to the cavity, relative phase noise of the two c.w. lasers is -113 dBc/Hz at 10 kHz offset frequency, when separated by one $FSR$ of the FP cavity. This is 64 dB lower than their free-running relative noise (Extended Fig. \ref{ExFig2}). 


For microwave generation, the remaining 25\% of the amplified microcomb output is directed to a fast photo detector (U2T XPDV2320R), generating a radio frequency signal of -7 dBm at $f_{\rm rep}$ (near 20 GHz). 
The microwave phase noise is shown in Fig. \ref{Fig1}f (measured using an R\&S FSWP). 
When scaled to a 10 GHz carrier (black curve), the SSB phase noise is -101 dBc/Hz at 100 Hz, -133 dBc/Hz at 1 kHz, and -152 dBc/Hz at 10 kHz, which is a record-low to our knowledge among photonic chip based platforms \cite{kudelin2023photonic,sun2023integrated,zhao2023all,yang2021dispersive,lucas2020ultralow,kwon2022ultrastable}. 
Fig. \ref{Fig1}g shows a representative radio frequency tone ($f_{\rm rep}$) under the 2P-OFD. 

\begin{figure*}[t!]
\begin{centering}
\includegraphics[width=170mm]{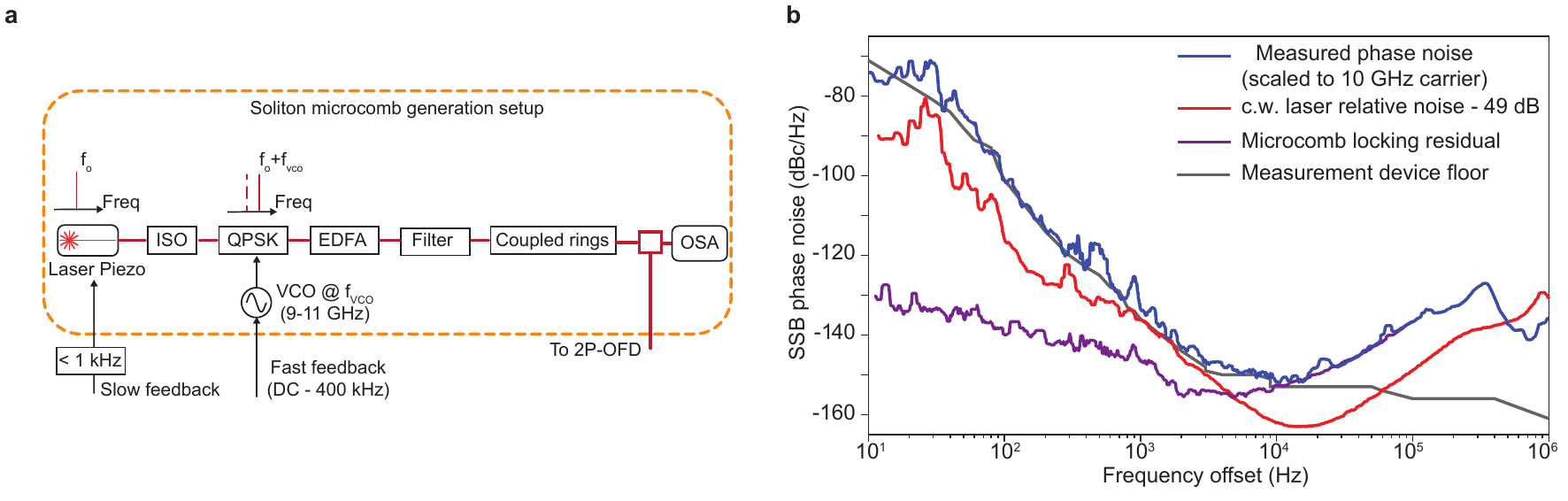}
\captionsetup{singlelinecheck=off, justification = RaggedRight}
\caption{{\bf 2P-OFD Details. }
\textbf{a}, Experimental setup for bright bright soliton microcomb generation. A c.w. laser (Orbits lightwave) near 1565 nm is isolated (ISO), frequency-shifted by a quadrature phase shift key modulator (QPSK), and amplified by an Erbium-doped fiber amplifier (EDFA). The light is then bandpass filtered to reduce the amplified spontaneous emission noise and coupled to the resonator chip using a lensed fiber. The coupled on-chip power is $\sim$150 mW. Most of the soliton power is routed to the 2P-OFD system (see text). A small portion is monitored by an optical spectrum analyzer. 
\textbf{b}, Phase noise spectra summary of the generated microwave signal in the 2P-OFD experiment. All noise levels are scaled to a 10 GHz carrier. 
}
\label{Fig3}
\end{centering}
\end{figure*}

\medskip

\noindent {\bf Dispersive-wave tunable microcomb.}
Concerning the microresonator design, the three rings share the same waveguide cross section, but the left (right) ring B (C) is $3\times 10^{-3}$ times larger (smaller) than the middle ring (ring A). The left (B) and right (C) rings are respectively coupled to the middle ring (A) by the evanescent field of neighbouring waveguides (2.4 $\mu$m coupling gap). The Si$_3$N$_4$ waveguides feature normal dispersion, but mode coupling enables generation of bright pulse pairs \cite{yuan2023soliton}. These pulse pairs form in certain spectral windows that can be tuned by electrical control of the ring temperatures using the heaters in Fig. \ref{Fig1}a. A two step tuning protocol is described in the extended data and Methods that allows configuration of the comb from any initial configuration. Likewise, this protocol permits higher-order dispersion control that tunes the dispersive wave wavelength. It is noted that after the differential tuning is completed, the resulting dispersion and microcomb spectrum are very stable. No change in dispersion or microcomb spectrum was observable over two months of measurements. 

Upon setup of the microcomb using this protocol, three dispersion bands are apparent as shown in Fig. \ref{Fig2}a, where measured (blue) and modeled (black) integrated dispersion $D_{\rm int}$ \cite{kippenberg2018dissipative} is plotted versus wavelength (see Methods) \cite{dorche2017extending,yuan2023soliton}.   
The upper band is used for microcomb generation and is pumped near $\lambda_{\rm o}=1565$ nm. Concerning the structure of these bands, there is no simple closed-form expression for the dispersion spectrum, but there are some universal features. The dispersion spectra of the uncoupled rings are indicated by the colored curves (red, blue and gray for the three rings, respectively). 
Where ring A crosses ring B and ring C curves (indicated by the two arrows) two bandgaps are opened upon introduction of ring coupling. 
Also note that ring B and ring C are able to indirectly couple to each other through their mutual interaction with ring A. This coupling creates another smaller bandgap. The magnitude of these gaps can be related to coupling strength and dispersion, and it discussed further in the Methods.



The dispersion curve of the 3CR system allows generation of double dispersive waves. 
These waves form at frequencies where soliton frequency components are nearly resonant with cavity modes (black arrows in Fig. \ref{Fig2}b). And controlled electrical (heater) tuning of the dispersive wave wavelength is used for matching to one of the c.w. lasers in the optical frequency division measurement. This capability greatly strengthens the beatnote SNR and enables flexible access to a wider range of c.w. lasers, whose wavelengths may not be widely tunable. Microcomb optical spectra showing different dispersive wave tunings are plotted in Fig. \ref{Fig2}c.  Tuning of the shorter wavelength dispersive wave by 4 nm and the longer wavelength dispersive wave by 8 nm is possible (see Extended Fig. \ref{ExFig1}). It is also noted that adjusting the pump laser-microresonator detuning allows tuning of the dispersive wave \cite{yang2016spatial}, but this provides only limited tuning range ($\sim$ 0.2 nm). As an aside, the bright soliton microcomb is triggered using the method described in the reference \cite{stone2018thermal}, which is also illustrated in Fig. \ref{Fig3}a.



\medskip

\noindent {\bf Discussion} Noise limits of the generated 20 GHz microwave tone are summarized in Fig. \ref{Fig3}b. The estimated phase noise limit from the relative phase noise of the two c.w. lasers is plotted in red. This is calculated by measuring the relative phase noise of the two lasers when they are PDH locked to the same mode family of the FP cavity, and when separated by one $FSR$ (4 GHz) of the FP cavity (Extended Fig. \ref{ExFig2}). 
It is assumed that the relative phase noise of the two c.w. laser does not change when they are separated by 3 THz in the 2P-OFD measurement (compared with the 4 GHz case). 
The phase noise is then scaled down by the 2P-OFD factor (49 dB to a 10 GHz carrier) to infer the noise limits in generating the 20 GHz microwave. In the current experiment only one of the two dispersive waves is used. In principle, use of both dispersive waves would increase the 2P-OFD factor by 6 dB to 55 dB. The locking residual of the microcomb servo is plotted in purple. Other possible limits in the phase noise include amplitude-to-phase conversion inside the photo detector \cite{Zhang2012, zang2018reduction}, quantum noise of the microcomb \cite{matsko2013timing}, amplified spontaneous emission noise from the EDFA, as well as photo detector shot noise\cite{quinlan2013exploiting}.

In summary, record low microwave phase noise levels have been demonstrated using a microcomb-based system. Key to this demonstration is the ability to electrically tune the spectral location of a dispersive wave. The coupled ring chip that generates the microcomb is fabricated at a CMOS foundry with high yield\cite{jin2021hertz}, and is thus suitable for mass production.  Moreover, the entire system could be potentially miniaturized for fieldable testing. The coupled ring structure has also been demonstrated under the self-injection lock mode and hybrid-integration with a III-V pump laser, thereby bypassing bulky fiber-optic components that may not be easy to integrate \cite{ji2023engineered}. 
The \FP cavity does not require operation in vacuum while being high-performance, reducing supporting hardware and operational complexity. And the two c.w. lasers  stabilized to the cavity can be readily miniaturized. The lasers are also stabilized without extra bulky fiber-optic components (\textit{e.g.} acoustic optic modulators) \cite{guo2022chip,kudelin2023photonic}. These combined features mean that the architecture simplifies the optical frequency division system and is potentially manufacturable and fieldable.


\noindent \textbf{Note added: } We would like to note two related papers that appeared during preparation of this manuscript \cite{jin2024microresonator,sun2024kerr}.

\bibliography{main.bib}

\newpage

\noindent\textbf{Methods}

\noindent \textbf{Dispersion modeling of the three-coupled-ring resonator} 
In this section, the theory that describes the dispersion spectrum of the 3-coupled-ring resonator is described.
Calculation of the dispersion is detailed in \cite{yuan2023soliton}. Briefly, a transfer matrix, $T$, is used to propagate a 3-component wave function through a round trip. 
\begin{widetext}
\begin{equation}
T=e^{i\omega \overline{L}/c}
\begin{pmatrix}
  e^{i 2\pi m(-\epsilon_1+\epsilon_2)}\cos(g_\mathrm{co}l_\mathrm{co}) &
i e^{-i4\pi m\epsilon_2}\cos(g_\mathrm{co}l_\mathrm{co})\sin(g_\mathrm{co} l_\mathrm{co}) &
- e^{i2\pi m (\epsilon_1+\epsilon_2)}\sin^2(g_\mathrm{co}l_\mathrm{co}) \\
i e^{i2\pi m(-\epsilon_1+\epsilon_2)}\sin(g_\mathrm{co}l_\mathrm{co}) &
  e^{-4i\pi m \epsilon_2}\cos^2(g_\mathrm{co} l_\mathrm{co}) & 
i e^{i2\pi m (\epsilon_1+\epsilon_2)}\cos(g_\mathrm{co}l_\mathrm{co})\sin(g_\mathrm{co}l_\mathrm{co}) \\
0 &
i e^{-i4\pi m\epsilon_2}\sin(g_\mathrm{co}l_\mathrm{co}) &
  e^{i2\pi m (\epsilon_1+\epsilon_2)}\cos(g_\mathrm{co}l_\mathrm{co}) 
\end{pmatrix}. 
\end{equation}
\end{widetext}
The resulting secular equation gives the eigenfrequencies, $\omega$, of the three mode families (see plot in Fig. \ref{Fig2}a),
where $m$ is mode number, 
$\epsilon_1 \equiv (L_{\rm B}-L_{\rm C}) / 2\overline{L}$, and $\epsilon_2 \equiv (L_{\rm B}+L_{\rm C}-2L_{\rm A})/6 \overline{L}$.
Here, $\overline{L} \equiv (n_{\rm wg,B}l_{\rm B}+n_{\rm wg,C}l_{\rm C})/2$ is the averaged optical path length of ring B and C (left and right rings), $L_{\rm i}\equiv n_{\rm wg, i}l_{\rm i}$ is the round-trip optical path length of any of the three rings (i=A,B,C), where  $n_{\rm wg, i}$ is the effective index of the waveguide that forms the ring (which can be tuned via the thermo-optic effect) and $l_{i}$ is the physical round trip length of an individual ring (which can be tuned via the thermo-elastic effect). 
$g_{\rm co}$ is the amplitude coupling strength per unit length between the neighbouring rings (ring A and B, ring A and C). 
$l_{\rm co}$ is the physical length of the coupling section.


The secular equation after a round trip can be simplified to a polynomial equation, 
\begin{widetext}
    \begin{equation}
    x^3
    -(e^{-2i\phi_2}\cos(g_\mathrm{co}L_\mathrm{co})+2e^{i\phi_2}\cos(\phi_1))\cos(g_\mathrm{co}l_\mathrm{co})x^2
    +(e^{2i\phi_2}\cos(g_\mathrm{co}l_\mathrm{co})+2e^{-i\phi_2}\cos(\phi_1))\cos(g_\mathrm{co}l_\mathrm{co})x
    -1=0, 
    \label{eqn:roots}
    \end{equation}
\end{widetext}
where $x \equiv e^{i\theta}$ and $\omega=\omega_{m}-\frac{D_{1}}{2\pi}\theta$, with $\omega_m=2\pi m c/\overline{L}$, $D_1/2\pi$ the average $FSR$ of the three rings, and $c$ the speed of light in vacuum. 
$\phi_1\equiv 2\pi m \epsilon_1$ and $\phi_2\equiv 2\pi m \epsilon_2$ are parameters that govern the dispersion spectrum. 
Three dispersion bands will be formed since eqn. (\ref{eqn:roots}) is a third order polynomial of $x$. 

The above eqn. (\ref{eqn:roots}) does not have a simple solution, but there are still features that can be inferred in terms of the dispersion spectrum. For example, $\phi_1=2\pi \epsilon_1 m_{\rm o}=2\pi N$ ($N$ is an integer) defines the mode number where the dispersion curves of the uncoupled ring B and ring C will intersect (Fig. \ref{Fig2}a). At this mode number the corresponding wavelength is $\lambda_{\rm o}$. For the device used in this study, $\epsilon_1\approx 3\times 10^{-3}$ by design and is measured to be within 1$\%$ of this value. $\epsilon_2$ is measured to be $\epsilon_2\sim 10^{-5}$ which is consistent with a design target close to zero. 
However, slight fabrication variances modify the dispersion curve and impair soliton mode locking at the pump wavelength.
To acquire dispersion bands that are favorable for soliton mode locking, heater tuning is applied to tune $\lambda_0$ ($m_{\rm o}$) and $\phi_2$. 
After the heater tuning, the dispersion is as shown in Fig. \ref{Fig2}a wherein the fitted parameters are $\phi_2=-0.36$, and $g_{\rm co} l_{\rm co}=0.9$ near $\lambda_{\rm o}=1568$ nm ($m_{\rm o} \approx$ 9567). Further details on this tuning procedure are given below.




\medskip 

\noindent \textbf{Differential heater tuning of the three-coupled-ring resonator.}
Based on the analysis in the previous section, the dispersion curve is determined by two parameters $\phi_1$ and $\phi_2$. 
Experimentally, ring B (or C) is thermally tuned to change $\phi_1$ and, in turn, $\lambda_{\rm o}$. Ring A is thermally tuned to change $\phi_2$ and, in turn, can be shown to tune the GVD parameter at $\lambda_{\rm o}$. Specifically, differential heating of ring A increases the curvature of the pumped band near $\lambda_{\rm o}$ (amount of anomalous dispersion), as in the inset of Fig. \ref{Fig2}c. These tuning steps are largely independent and enable a 2-step dispersion tuning protocol described in Extended Fig. \ref{ExFig1}a. 
In the first step, ring B is tuned ($\phi_1$ is tuned) such that $\lambda_{\rm o}$ is tuned close to the pump wavelength (1565 nm). In the second step, ring A is tuned ($\phi_2$ is tuned) with $\phi_1$ unchanged. The resulting dispersion is shown in the lower panel of Extended Fig. \ref{ExFig1}a (also in Fig. \ref{Fig2}b). 

The dispersion tuning is efficient, requiring only a moderate amount of actual temperature tuning. Tuning of $\lambda_{\rm o}$ benefits from the Vernier effect as described in the reference \cite{ji2023integrated}, and experimentally $\sim$ 10 $^{\rm o}$C of differential temperature tuning is sufficient to tune the pump wavelength $\lambda_{\rm o}$ across the optical C band. 
The local curvature of the dispersion bands (GVD parameter) near $\lambda_{\rm o}$ is determined by differential thermal tuning in ring A. 
The tuning of uncoupled mode resonance in ring A by one $FSR$ (20 GHz)
will access all the possible dispersion configurations. This corresponds to tuning of $L_{\rm A}$ by $\overline{L}/m$. 
With a large mode number $m\sim 10^4$ in the optical C band, this corresponds to $< 10$ $^{\rm o}$C of differential temperature tuning.

\medskip 
\noindent \textbf{High-finesse, vacuum-free and miniaturized \FP cavity} The cavity is made of high purity fused silica with high reflectivity coatings on both sides to form resonance.
It is cylindrical in shape, 25.4 mm in length and 15 mm in diameter; see the photograph of Fig. 1c. Thedielectric coatings (SiO$_2$/Ta$_2$O$_5$) at 1550 nm are deposited on both surfaces (plano-convex with 1 meter radius of curvature). Since the cavity is solid silica, a vacuum chamber and associated equipment are not required. By using a c.w. laser to sweep across the cavity resonance, the reflection of the cavity reveals the cavity linewidth to be 24 kHz (quality factor 8.2 billion); see the measurement and fitting in inset of Fig. \ref{Fig1}c. The vibration sensitivity of the cavity is designed and measured to be $10^{-10}/g$ by using the vibration-insensitive scheme demonstrated in ref~\cite{zhang2020ultranarrow}. The \FP cavity is installed in a multi-layer thermal shield, providing insulation ($\approx$ 20 mins time constant) to damp the impact of ambient temperature fluctuation. 

By Pound-Drever-Hall (PDH) locking individual lasers (Toptica laser and the RIO lasers) to neighboring fundamental modes of the cavity~\cite{drever1983laser} and then detecting the beatnote of the two lasers with a fast photodetector (Thorlabs DXM30AF), a 4 GHz microwave tone (one cavity $FSR$) is generated. This tone both with and without PDH lock is measured using an R\&S FSWP signal analyzer (see Extended Data Fig. \ref{ExFig2}). The phase noise spectra show a large noise reduction from the cavity locking. The red curve in Extended Data Fig. \ref{ExFig2} is reproduced in \ref{Fig3}b for comparison to the 2P-OFD microwave phase noise spectrum.  Considering the noise reduction due to 2P-OFD, which is 49 dB (when scaled to a 10 GHz carrier), the projected microwave phase noise is approximately $>$10 dB lower than the measured phase noise of the comb repetition rate (blue). 


\medskip 

\noindent \textbf{Details of the 3CR resonator. }
For the device used in the study, the intrinsic $Q$ factor of the pumped mode is around 100 million, and the experimental threshold power for parametric oscillation is around 3 mW. 
Also, the $FSR$ is 19.869 GHz and the local GVD parameter is $D_2/2\pi\approx 50$ kHz. Other parameters can be found in the reference \cite{ji2023engineered}. 

\vspace{3 mm}

\noindent \textbf{Data Availability}
The data that supports the plots within this paper and other findings of this study are available from the corresponding author upon reasonable request.

\vspace{1 mm}

\noindent \textbf{Code Availability}
The codes that support findings of this study are available from the corresponding author upon reasonable request.

\vspace{1 mm}
\noindent \textbf{Acknowledgments}
The authors thank Igor Kudelin, Frank Quinlan and Scott Diddams at NIST,  Shuman Sun and Xu Yi at University of Virginia for fruitful discussion, as well as Maodong Gao, Jinhao Ge, Yu Yan and Zhiquan Yuan at Caltech for experimental assistance.
This work is supported by the NASA (Award number NASA.23M0239), Defense Advanced Research Projects Agency GRYPHON (HR0011-22-2-0009)and the Kavli Nanoscience Institute at Caltech. 
The research reported here performed by W.Z. V.I. and A.M. was carried out at the Jet Propulsion Laboratory at the California Institute of Technology, under a contract with the National Aeronautics and Space Administration (80NM0018D0004). 
The content of the information does not necessarily reflect the position or the policy of the federal government, and no official endorsement should be inferred.
\vspace{1 mm}

\noindent\textbf{Author Contributions} Concepts were developed by Q.-X.J. W.Z., P.L., W.J., J.B., A.M., K.V. . 
W.J. designed and fabricated the 3CR resonator with inputs from J.G. J.P. L.W. Q.-X.J.  A.F. M.F. J.B..
W.Z. designed and built the FP cavity subsystem with assistance from Q.-X.J..
Measurements and modeling were performed by Q.-X.J. W.Z. P.L.. 
All authors contributed to the writing of the manuscript. 
The project was supervised by W.Z. A.M. J.B. K.V. 
\vspace{1 mm}

\noindent \textbf{Competing Interests} The authors declare no competing interests.

\vspace{1 mm}

\noindent \textbf{Author Information} Correspondence and requests for materials should be addressed to W.Z. (wei.zhang@jpl.nasa.gov) or K.V. (vahala@caltech.edu).

\clearpage
\onecolumngrid
\renewcommand{\figurename}{\bf Extended Data Fig.}
\setcounter{figure}{0}

\newpage

\begin{figure*}[t!]
\begin{centering}
\includegraphics[width=\linewidth]{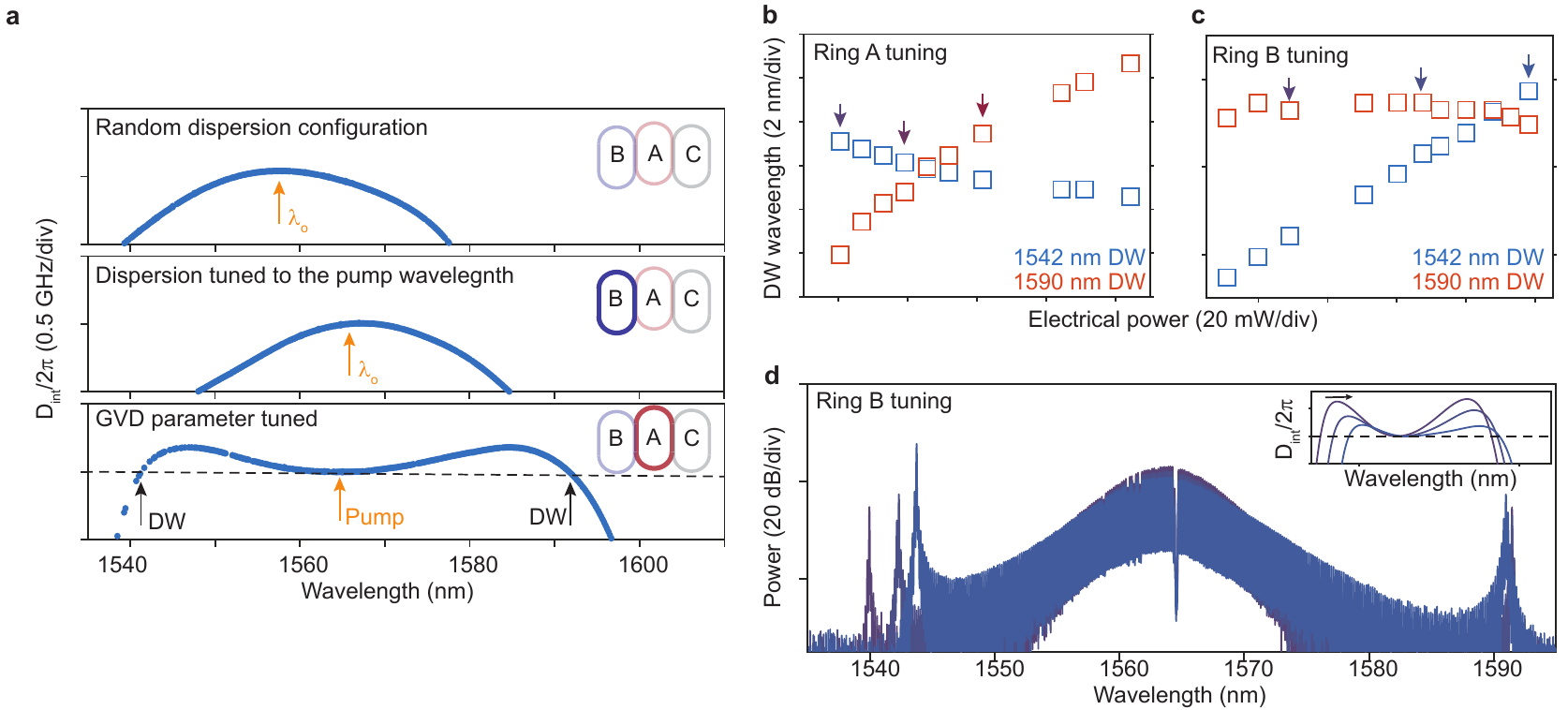}
\captionsetup{singlelinecheck=off, justification = RaggedRight}
\caption{{\bf Dispersion tuning and dispersive wave tuning in the bright soliton microcomb using electrical heaters.}
\textbf{a}, Upper panel: measured dispersion spectrum of the 3-coupled-ring (3CR) resonator with random heater tuning. 
Middle panel: measured dispersion spectrum of the 3CR after tuning $\lambda_{\rm o}$ closer to the pump wavelength (1565 nm) by heating ring B.
Lower panel: measured dispersion spectrum of the 3CR after the local GVD parameter is tuned by heating ring A.  
\textbf{b,c}, Measurement of dispersive wave tuning when ring A (B) is fine-tuned. The 1542 nm dispersive wave is plotted in blue, while the 1590 nm dispersive wave is plotted in red. The corresponding data points in Fig. \ref{Fig2}c and panel d are indicated by the arrows (matched by the colours). 
\textbf{d}, Measured optical spectra of the bright soliton microcomb with three different dispersive wave tunings. 
Inset: theoretical dispersion spectra when ring B is tuned. The arrow indicates the change of the dispersion curves when ring B is heated.
}
\label{ExFig1}
\end{centering}
\end{figure*}

\begin{figure}[t!]
\begin{centering}
\includegraphics[width=\linewidth]{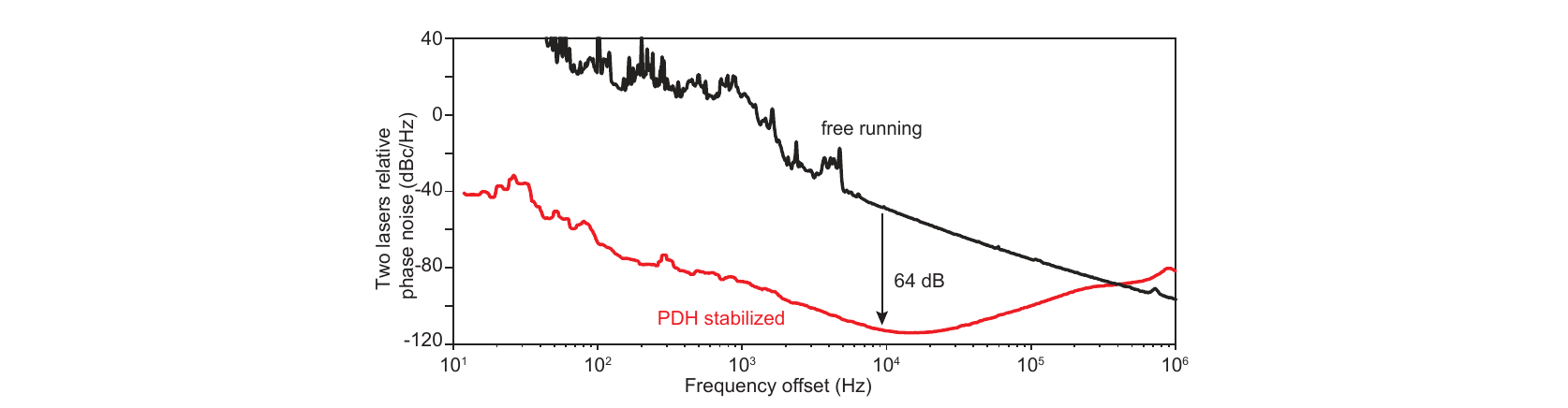}
\captionsetup{singlelinecheck=no, justification = RaggedRight}
\caption{
Relative phase noise of the two c.w. lasers, when both free-running (black) and PDH stabilized to the FP cavity (red).  
}
\label{ExFig2}
\end{centering}
\end{figure}

\begin{figure*}[t!]
\begin{centering}
\includegraphics[width=170mm]{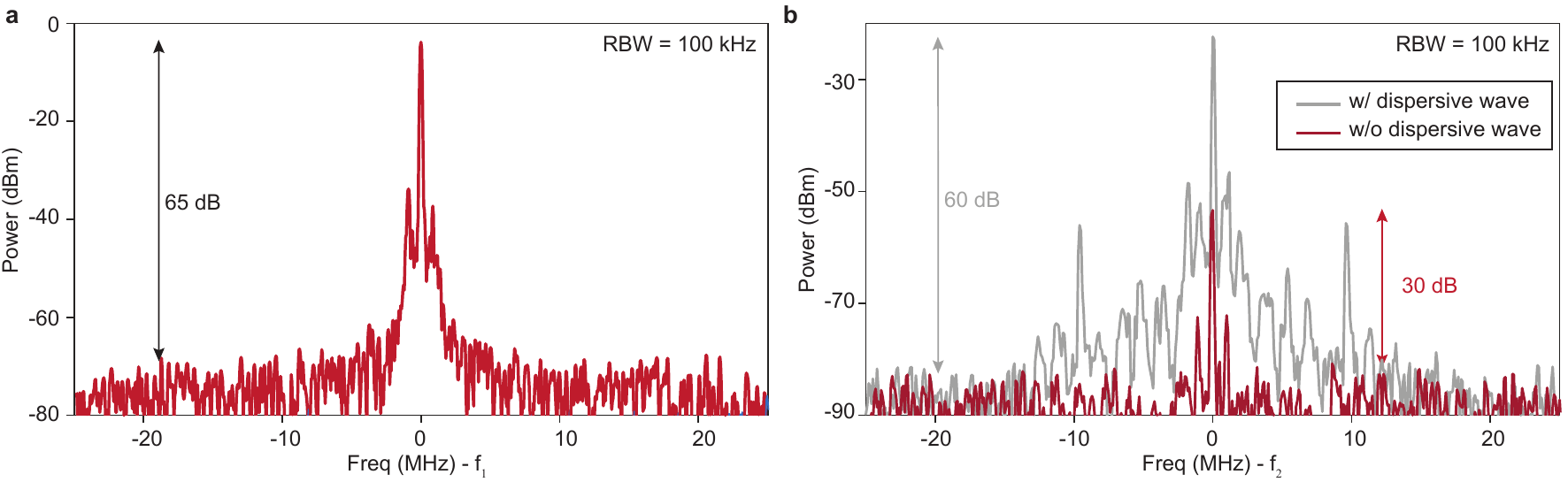}
\captionsetup{singlelinecheck=off, justification = RaggedRight}
\caption{{\bf Measured beatnotes $f_1$ and $f_2$ in Fig. \ref{Fig1} with and without dispersive-wave enhancement. }
\textbf{a}, Measured beatnote between the stabilized c.w. laser ($\nu_1$) and comb line ($\nu_m$), where $f_1$ is 507.4 MHz. 
An SNR of 65 dB is measured with resolution bandwidth of 100 kHz. 
\textbf{b}, Measured beanote between the stabilized c.w. laser ($\nu_2$) and the dispersive wave ($\nu_n$). 
An SNR of 60 dB is measured with resolution bandwidth of 100 kHz. 
$f_2$ is 743.6 MHz for the case with dispersive wave, and $f_2 =979.9$ MHz for the case without the dispersive wave. 
A 30 dB improvement in SNR is demonstrated with the dispersive wave. 
}
\label{ExFig4}
\end{centering}
\end{figure*}

\clearpage
\newpage

\end{document}